\documentstyle[pra,aps,preprint]{revtex}
\begin{document}
\draft
\title{Quantum Electrodynamical Density-matrix Functional Theory and Group-theoretical Consideration of its Solution}
\author{Tadafumi Ohsaku\footnote{Corresponding author; Tadafumi Ohsaku, tadafumi@fuji.phys.wani.osaka-u.ac.jp}}
\address{Department of Physics, and Department of Chemistry, Graduate School of Science, Osaka University, Toyonaka, Osaka, Japan, and Research Center for Nuclear Physics (RCNP), Osaka University, Ibaraki, Osaka, Japan}
\author{Shusuke Yamanaka, Daisuke Yamaki and Kizashi Yamaguchi}
\address{Department of Chemistry, Graduate School of Science, Osaka University, Toyonaka, Osaka, Japan}

\maketitle

\begin{abstract}

For studying the group theoretical classification of the solutions of the density functional theory in relativistic framework, we propose quantum electrodynamical density-matrix functional theory ( QED-DMFT ). QED-DMFT gives the energy as a functional of a local one-body $4\times4$ matrix $Q(x)\equiv -\langle\psi(x)\bar{\psi}(x)\rangle$, where $\psi$ and $\bar{\psi}$ are 4-component Dirac field and its Dirac conjugate, respectively. We examine some characters of QED-DMFT. After these preparations, by using $Q(x)$, we classify the solutions of QED-DMFT under $O(3)$ rotation, time reversal and spatial inversion. The behavior of $Q(x)$ under nonrelativistic and ultrarelativistic limits are also presented. Finally, we give plans for several extensions and applications of QED-DMFT.

\end{abstract}

keywords; density functional theory, QED, QED-DMFT, QED-TDDMFT, QED-CAS-DMFT, group theory, symmetry and broken symmetry, heavy elements.

\section{Introduction}

The density matrix was first introduced by Dirac~[1], and after it was generalized by Husimi~[2] and L\"{o}wdin~[3] ( the reduced density matrix ). This was recognized as a very useful tool to investigate interacting many-body theory, and it gave various important concepts until now. The definition of the one-particle density matrix is given as
\begin{eqnarray}
\gamma(x,x')_{\alpha\beta}=\langle\psi^{\dagger}_{\beta}(x')\psi_{\alpha}(x)\rangle. 
\end{eqnarray}
Here $\psi$ and $\psi^{\dagger}$ are 2-component electron field and its hermitian conjugate, respectively. On the other hand, the density functional theory ( DFT ) was first given by Hohenberg-Kohn-Sham~[4], and it was developed by many researchers. People applied DFT to various many-fermion systems, not only many-electron systems like atoms, molecules and solids, but also nuclear systems. Today, DFT is well accepted, and used to investigate various physical and chemical characters of systems~[5$\sim$7].

The symmetry aspect of DFT was also discussed by using the one-particle local $2\times 2$ matrix density:
\begin{eqnarray}
\rho(x)_{\alpha\beta}=\langle\psi^{\dagger}_{\beta}(x)\psi_{\alpha}(x)\rangle. 
\end{eqnarray}
By using (2), G\"{o}rling examined the symmetry of the Hohenberg-Kohn theorem in detail~[8]. Weiner and Trickey investigated the group theoretical classification of nonrelativistic DFT solutions~[9], based on the work of Fukutome in which the classification for generalized Hartree-Fock ( GHF ) theory was presented~[10]. Yamanaka et al.~[11] applied the generalized spin orbital ( GSO ) DFT to various molecular systems and studied the symmetry breaking phenomena in DFT solutions. We should mention also, it was demonstrated that the Hohenberg-Kohn theorem varid for one-particle density matrix functional $E[\gamma(x,x')]$~[12$\sim$15]. DFT for time-dependent phenomena was also investigated ( the time-dependent density functional theory ( TDDFT ) ). Runge and Gross examined the Hohenberg-Kohn theorem for time-dependent external potential ( the Runge-Gross theorem )~[16]. TDDFT was applied to estimation of electronic excitation or susceptibilities~[17]. 

By the way, the first generalization of relativistic Hohenberg-Kohn theorem was introduced by Rajagopal-Callaway~[18], and several attempts were made to investigate relativistic density functional theory ( RDFT )~[19$\sim$24]. Based on these studies, an important review of RDFT was presented by Eschrig et al.~[25]. Today, the Hohenberg-Kohn-Sham scheme is well established for RDFT~[26$\sim$28]. Relativistic time-dependent density functional theory ( RTDDFT ) was also investigated~[29,30]. 

In this note we investigate quantum electrodynamical density-matrix functional theory ( QED-DMFT ), and its group theoretical aspects. Recently, we  constructed the theory of quantum electrodynamical self-consistent fields ( QED-SCF )~[31]. We derived the time-dependent Hartree-Fock ( QED-TDHF ) theory, Hartree-Fock condition and random phase approximation ( QED-RPA ). We also presented the group theoretical classification of quantum electrodynamical Hartree-Fock ( QED-HF ) solutions~[32]. Based on our previous results, in this work we apply the group-theoretical method to QED-DMFT. Our theory can become a basis or a starting point to investigate the phenomena of symmetry breaking in many-body system~[9$\sim$11,33] under strong relativistic effects, by using DFT framework. 
In the work of Yamanaka et al.~[11], the results indicate that the GSO treatment of DFT is very important and usuful for description of some polyradical clusters. Therefore, it is interesting for us to extend their treatment to the relativistic framework, and study the phenomena of symmetry breaking in heavy elements, analogous with the nonrelativistic GHF theory~[10]. 

This paper is organized as follows. In Sec. II, we introduce QED-DMFT. Here, ``matrix'' means a matrix in spinor space. In fact, we do not use a non-local one-body matrix like $\gamma(x,x')_{\alpha\beta}$ in nonrelativistic case, rather we intend to discuss a local one-body matrix like $\rho(x)_{\alpha\beta}$ in nonrelativistic case for obtaining simple classification under group operations. The relation between various available DFT and QED-DMFT is also discussed. Quantum electrodynamical time-dependent density-matrix functional theory ( QED-TDDMFT ) is also investigated. In Sec. III, the group theoretical classification of QED-DMFT solutions is discussed. The results can be applied to elucidate the phenomena of broken symmetry, especially in atomic systems under relativistic effects. Finally in Sec. IV, possible extensions and applications of QED-DMFT are discussed. Conclusion with summary of this work is also presented.

\section{QED-DMFT}

\subsection{Introduction for QED-DMFT}

In this subsection, we introduce QED-DMFT. First, we give a brief summary of QED. We start from the following Lagrangian:
\begin{eqnarray}
{\cal L}(x) &=& -\frac{1}{4}F_{\mu\nu}(x)F^{\mu\nu}(x) + \bar{\psi}(x)(i\gamma^{\mu}\partial_{\mu}-m_{0})\psi(x) +e\bar{\psi}(x)\gamma^{\mu}\psi(x)(A_{\mu}(x)+A^{ext}_{\mu}(x)).
\end{eqnarray}
Here, $A_{\mu}$ is a ( virtual ) photon field which describes the interaction
between fermions, and $A^{ext}_{\mu}$ is a classical external field. $F_{\mu\nu}(x)=\partial_{\mu}A_{\nu}(x)-\partial_{\nu}A_{\mu}(x)$. $m_{0}$ is the bare mass of fermion, $\psi(x)$ is the Dirac field, $\bar{\psi}(x)$ is its Dirac conjugate. $x\equiv(x_{0},{\bf x})$ indicates time-space coordinate. The $\gamma$-matrices are defined by ( under the standard representation )
\begin{equation}
\gamma^{0}=\left(
\begin{array}{cc}
I & 0 \\
0 & -I 
\end{array}
\right), \qquad \gamma^{i}=\left(
\begin{array}{cc}
0 & \sigma^{i} \\
-\sigma^{i} & 0
\end{array}
\right),
\end{equation}
where $I$ is the unit matrix, and $\sigma_{i}$ ( i=1,2,3 ) is the Pauli matrix:
\begin{equation}
\sigma^{1}=\left(
\begin{array}{cc}
0 & 1 \\
1 & 0 
\end{array}
\right), \qquad \sigma^{2}=\left(
\begin{array}{cc}
0 & -i \\
i & 0 
\end{array}
\right), \qquad \sigma^{3}=\left(
\begin{array}{cc}
1 & 0 \\
0 & -1 
\end{array}
\right).
\end{equation}
The Hamiltonian for a QED system becomes
\begin{eqnarray}
\hat{H} &=& \int d^{3}x\hat{\bar{\psi}}(x)(-i\vec{\gamma}\cdot\nabla+m_{0})\hat{\psi}(x) \nonumber \\
  & & +\int d^{3}x\hat{\bar{\psi}}(x)\gamma^{\mu}\hat{\psi}(x)(\hat{A}_{\mu}(x)+A^{ext}_{\mu}(x))+\hat{H}_{em}, 
\end{eqnarray}
where, $A^{ext}_{\mu}$ is a classical field, while others are quantized. 
$\hat{H}_{em}$ is a Hamiltonian for electromagnetic field. If we descibe the interaction between fermions by using a photon propagator, we get
\begin{eqnarray}
\hat{H}' &=& \int d^{3}x\hat{\bar{\psi}}(x)(-i\vec{\gamma}\cdot\nabla+m_{0}+\gamma^{\mu}A^{ext}_{\mu}(x))\hat{\psi}(x) \nonumber \\
  & & +\frac{e^{2}}{2}\int d^{3}x\int d^{4}y \hat{\bar{\psi}}(x)\gamma^{\mu}\hat{\psi}(x)D_{\mu\nu}(x-y)\hat{\bar{\psi}}(y)\gamma^{\nu}\hat{\psi}(y). 
\end{eqnarray}
Here, $D_{\mu\nu}(x-y)$ is a full photon propagator.
Quantized fermion-field operators $\hat{\psi}(x)$ and $\hat{\bar{\psi}}(x)$ can be expanded by one-particle state
functions as
\begin{eqnarray}
\hat{\psi}(x) &=& \sum_{i}(\psi^{(+)}_{i}(x)\hat{a}_{i}+\psi^{(-)}_{i}(x)\hat{b}^{\dagger}_{i}), \\
\hat{\bar{\psi}}(x) &=& \sum_{i}(\bar{\psi}^{(+)}_{i}(x)\hat{a}^{\dagger}_{i}+\bar{\psi}^{(-)}_{i}(x)\hat{b}_{i}),  
\end{eqnarray}
where $+(-)$ means the electron ( positron ) state, $i$ denotes quantum numbers for one-particle state, $\hat{a}^{\dagger}_{i}(\hat{a}_{i})$ is electron creation ( annihilation ) operator, $\hat{b}^{\dagger}_{i}(\hat{b}_{i})$ is positron creation ( annihilation ) operator, and $\psi^{(\pm)}_{i}$ is the four-component bispinor given like
\begin{eqnarray}
\psi^{(\pm)}_{i}=(f^{(\pm)}_{i(1)},f^{(\pm)}_{i(2)},f^{(\pm)}_{i(3)},f^{(\pm)}_{i(4)})^{T},
\end{eqnarray} 
where, $T$ indicates the transposition of a matrix. 

After these preliminaries given above, we investigate the QED-DMFT. 
For our aim, first we introduce the one-particle local density matrix:
\begin{equation}
Q(x)_{4\times4} = -\langle\psi(x)\bar{\psi}(x)\rangle_{4\times4}.
\end{equation}  
Here $\psi$ and $\bar{\psi}$ are usual Dirac field, and they are 4-component bispinors. Thus our density matrix is 4$\times$4 matrix, as denoted above.
We can expand the 4$\times$4 density matrix into the 16-dimensional complete set of $\gamma$-matrices:
\begin{eqnarray}
Q_{4\times4} &=& \sum^{16}_{A=1}\Gamma_{A}Q^{A} = Q^{S}I+Q^{V}_{\mu}\gamma^{\mu}+Q^{T}_{\mu\nu}\sigma^{\mu\nu}+Q^{A}_{\mu}\gamma_{5}\gamma^{\mu}+Q^{P}i\gamma_{5}, \\
Q^{A} &=& \frac{1}{4}{\rm tr}\Gamma_{A}Q = \frac{1}{4}\langle\bar{\psi}\Gamma_{A}\psi\rangle. 
\end{eqnarray}
In this expansion, we take a convention as $S$ denotes the scalar, $V$ denotes the vector, $T$ denotes the 2-rank antisymmetric tensor, $A$ denotes the axial vector and $P$ denotes the pseudoscalar. $I$ is the $4\times4$ unit-matrix, $\gamma^{\mu}$ is usual Dirac gamma matrix, $\sigma^{\mu\nu}$ is defined as $\sigma^{\mu\nu}=\frac{i}{2}[\gamma^{\mu},\gamma^{\nu}]$, and $\gamma_{5}=i\gamma^{0}\gamma^{1}\gamma^{2}\gamma^{3}$. Hence we obtain the Lorentz structure in our density matrix $Q$. Each component of the expansion in (12) is given as follows: $Q^{S}=\frac{1}{4}{\rm tr}Q=\frac{1}{4}\langle\bar{\psi}\psi\rangle$, $Q^{V}_{\mu}=\frac{1}{4}{\rm tr}\gamma^{\mu}Q=\frac{1}{4}\langle\bar{\psi}\gamma^{\mu}\psi\rangle$, $Q^{T}_{\mu\nu}=\frac{1}{4}{\rm tr}\sigma^{\mu\nu}Q=\frac{1}{4}\langle\bar{\psi}\sigma^{\mu\nu}\psi\rangle$, $Q^{A}_{\mu}=\frac{1}{4}{\rm tr}\gamma_{5}\gamma^{\mu}Q=\frac{1}{4}\langle\bar{\psi}\gamma_{5}\gamma^{\mu}\psi\rangle$ and $Q^{P}=\frac{1}{4}{\rm tr}i\gamma_{5}Q=\frac{1}{4}\langle\bar{\psi}i\gamma_{5}\psi\rangle$. Therefore, $Q(x)$ includes a larger amount of informations than 4-current $j_{\mu}(x)=-e\langle\bar{\psi}(x)\gamma^{\mu}\psi(x)\rangle$. It is clear from the discussion given above, if $Q(x)$ is given, then $j_{\mu}(x)$ and also other components can be calculated.

The Rayleigh-Ritz variation principle of QED-DMFT should be given rather straightforward extension of available theory, like the situation of the generalization of the Hohenberg-Kohn theorem to 4-current RDFT~[7,14,18,25,26]. The exact energy functional of N-representable $Q(x)$ for ground state of a QED system ( Strictly spearking, the particle number $N$ is not conserved in QED, but the total charge is conserved, as denoted in (15). ) is given in the following form:
\begin{eqnarray}
E[Q(x)] &=& \langle \Psi^{Q}|\hat{H}'|\Psi^{Q}\rangle \nonumber \\ 
        &=& F[Q(x)] + \int d^{3}x j^{\mu}({\bf x})A^{ext}_{\mu}({\bf x}).
\end{eqnarray}
Here, we use the Hamiltonian $\hat{H}'$ given in (7).
Now $A^{ext}_{\mu}({\bf x})$ is a static external field. Here we only consider a stationary problem. 
Variational search of the minimum point of the functional should be performed under constraint:
\begin{eqnarray}
\partial_{\mu}j^{\mu}(x)=0, \qquad \int d^{3}x j_{0}({\bf x}) = const. 
\end{eqnarray}
First equation is arised from the gauge invariance of the theory derived by N\"{o}ther theorem, while second one expresses the charge conservation. The fundamental theorems of DFT proved by Levy are constructed by two points~[14]: (i) The variation principle can be applied to the ground state energy functional $E[\rho(x)]$ or $E[\gamma(x, x')]$, where $\rho(x)$ is the one-particle density while $\gamma(x,x')$ is the one-particle ( so called first-order ) density matrix. (ii) The ground state of a system can be represented by $\rho(x)$ or $\gamma(x, x')$ for ground state ( $\rho_{GS}(x)$ or $\gamma_{GS}(x, x')$ ). It is clear from his logic, the theorems do not distinguish between nonrelativistic case and relativistic case, and can be applied to relativistic case. Therefore, we argue that our functional $E[Q(x)]$ will obey the variation principle and $Q(x)$ for the ground state ( $Q_{GS}(x)$ ) represents the ground state of a QED system. Thus
\begin{eqnarray}
F[Q] &=& \min_{\Psi\to Q}\langle\Psi|\hat{T}+\hat{V}_{f-f}|\Psi\rangle, \\
E[Q] &=& \min_{Q}\Bigl\{F[Q]+\int d^{3}x j^{\mu}({\bf x})A_{\mu}({\bf x})\bigg|\int d^{3}x j_{0}({\bf x})=const. \Bigr\},
\end{eqnarray}
where $\hat{T}$ is the kinetic energy while $\hat{V}_{f-f}$ is the interaction between fermions, presented in the Hamiltonian (7).

Next step of our plan is the investigation into the Kohn-Sham scheme in our theory. The energy functional is devided into the following form:
\begin{eqnarray}
E[Q(x)] &=& T_{s}[Q(x)]+\int d^{3}x j^{\mu}({\bf x})A^{ext}_{\mu}({\bf x})+\frac{1}{2}\int d^{3}x \int d^{3}x'\frac{j^{\mu}({\bf x})j_{\mu}({\bf x}')}{|{\bf x}-{\bf x}'|}+E_{XC}[Q(x)]. 
\end{eqnarray}
Here, $T_{s}[Q]$ is the one-particle kinetic energy functional
\begin{eqnarray}
T_{s}[Q]=\int d^{3}x\sum_{-m<\epsilon_{n}\le\epsilon_{F}}\bar{\psi}_{n}(-i\vec{\gamma}\cdot\nabla+m_{0})\psi_{n},
\end{eqnarray}
and the 4-current $j_{\mu}({\bf x})$ is given as
\begin{eqnarray}
j_{\mu}({\bf x})=-e\sum_{-m<\epsilon_{n}\le\epsilon_{F}}\bar{\psi}_{n}({\bf x})\gamma_{\mu}\psi_{n}({\bf x}),
\end{eqnarray}
where, $\psi_{n}$ is the Kohn-Sham one-body function in our theory. $E_{XC}[Q(x)]$ is the exchange-correlation energy functional in our theory. The definition of $E_{XC}[Q(x)]$ is given as
\begin{eqnarray}
E_{XC}[Q(x)] &=& F[Q(x)]-T_{s}[Q(x)]-\frac{1}{2}\int d^{3}x \int d^{3}x'\frac{j^{\mu}({\bf x})j_{\mu}({\bf x}')}{|{\bf x}-{\bf x}'|}. 
\end{eqnarray}
The third term of the right hand side of (18) is a Hartree-like term. This term arises if we replace $D_{\mu\nu}(x-y)$ of (7) to the 0-th order photon propagator of the Feynman gauge:
\begin{eqnarray}
D^{(0)}_{\mu\nu}(x-y)=\int\frac{dk_{0}}{2\pi}e^{-ik_{0}(x_{0}-y_{0})}g_{\mu\nu}\frac{e^{i|k_{0}||{\bf x}-{\bf y}|}}{4\pi|{\bf x}-{\bf y}|}.
\end{eqnarray}
Here $g_{\mu\nu}$ is usual metric tensor and defined as $g_{\mu\nu}={\rm diag}(1,-1,-1,-1)$. Therefore, the third term of (18) includes gauge parameter. $E_{XC}$ has to include contributions to compensate the gauge dependence of the Hartree-like term, for retaining the gauge invariance of the theory. From the variation with respect to the one-particle function, we obtain the Dirac-Kohn-Sham equation in our theory:
\begin{eqnarray}
(-i\vec{\gamma}\cdot\nabla+m_{0}+\sum^{16}_{A=1}\Gamma^{A}v_{A}({\bf x}))\psi_{n}(x) = \epsilon_{n}\gamma^{0}\psi_{n}(x), \\
v^{S}({\bf x}) = \frac{\delta E_{XC}[Q(x)]}{\delta Q^{S}(x)} = \delta m +\alpha({\bf x}), \\
v^{V}_{\mu}({\bf x}) = A^{ext}_{\mu}({\bf x})+\int d^{3}x'\frac{j_{\mu}({\bf x}')}{|{\bf x}-{\bf x}'|}+\frac{\delta E_{XC}[Q(x)]}{\delta Q^{V}_{\mu}(x)}, \\
v^{T}_{\mu\nu}({\bf x}) = \frac{\delta E_{XC}[Q(x)]}{\delta Q^{T}_{\mu\nu}(x)}, \\
v^{A}_{\mu}({\bf x}) = \frac{\delta E_{XC}[Q(x)]}{\delta Q^{A}_{\mu}(x)}, \\
v^{P}({\bf x}) = \frac{\delta E_{XC}[Q(x)]}{\delta Q^{P}(x)}.     
\end{eqnarray}
Because our energy functional depends on the matrix $Q(x)$, generally we obtain 16-component fictitious potentials. From the symmetry consideration, it is clear that, scalar potential $v^{S}$ gives the mass correction $\delta m$, which will arises from radiative corrections. $v^{V}_{\mu}$ appears also in usual 4-current RDFT~[7,26]. In fact, QED-DMFT is a non-perturbative method in QED. It is a famous fact that, QED becomes finite if its calculation fulfills the requirement of the gauge invariance~[34] ( though we should employ gauge-invariant regularization and renormalization ). Therefore, if we can perform a calculation under retaining gauge invariance exactly, we will obtain a finite value for QED-DMFT.

\subsection{Relations between various DFT Schemes}

Now, we discuss the relation between various DFT, as depicted in Fig. 1. If we take into account only the components of 4-current $j_{\mu}$ in the expansion (12) of $Q(x)$ for QED-DMFT energy functional, we obtain quantum electrodynamical current density functional theory ( QED-CDFT )~[7,25,26]. Next, under performing the Gordon decomposition to spatial components of current ${\bf j}({\bf x})={\bf I}+\frac{1}{m}\nabla\times{\bf S}$ ( here ${\bf I}=\frac{i}{2m}\{\bar{\psi}\nabla\psi-(\nabla\bar{\psi})\psi\}$ is the orbital current density while ${\bf S}=\frac{1}{2}\bar{\psi}\gamma_{5}\gamma^{0}\vec{\gamma}\psi$ is the spin density ), and then completely omit ${\bf I}$, we obtain relativistic spin density functional theory ( RSDFT )~[25,28]. Next, neglect the contribution of ${\bf S}$ for RSDFT, then we obtain relativistic density functional theory ( RDFT ). Performing some nonrelativistic reduction, especially only take into account the contribution of the large component to RDFT, and neglect all relativistic corrections ( spin-orbit, Darwin, mass-velocity etc. ), then we obtain nonrelativistic density functional theory ( NRDFT )~[5$\sim$7]. Generalizing the NRDFT to spin polarized system, we get nonrelativistic spin density functional theory ( NRSDFT )~[5$\sim$7]. Add the interaction between orbital current and external vector potential ${\bf A}^{ext}$ to NRSDFT, we obtain nonrelativistic current density functional theory ( NRCDFT )~[35].

\subsection{QED-TDDMFT}

Extension of QED-DMFT to the time-dependent region can also be considered. 
Starting point here is the action integral for density functional, like the QED-TDHF~[31]:
\begin{eqnarray}
S[Q] = \int^{t_{f}}_{t_{i}}dt\langle\Psi^{Q}(t)|i\frac{\partial}{\partial t}-\hat{H}'|\Psi^{Q}(t)\rangle.
\end{eqnarray}
Here matrix $Q(x)=Q(x_{0},{\bf x})$, depends on time $x_{0}$. $H'$ is given as (7). We take the Dirac-Frenkel variation principle to produce a stationary point at time-dependent density $Q(x)$~[17,31]:
\begin{eqnarray}
\delta S[Q] = \delta \int^{t_{f}}_{t_{i}}dt\langle\Psi^{Q}(t)|i\frac{\partial}{\partial t}-\hat{H}'|\Psi^{Q}(t)\rangle =0.
\end{eqnarray}
Therefore we obtain the Euler equation:
\begin{eqnarray}
\frac{\delta S[Q]}{\delta Q_{A}(x_{0},{\bf x})}=0, \qquad A=1\sim16.
\end{eqnarray}
The Hohenberg-Kohn theorem in time-dependent potential was discussed by Runge-Gross~[16,17] for nonrelativistic case, and it was extended by Rajagopal~[30] for relativistic case, in detail. In fact, we simply assume that our functional of $Q$ can be used to discuss variation principle parallel with their results. 

The time-dependent Dirac-Kohn-Sham equation in our theory becomes  
\begin{eqnarray}
(i\gamma^{\mu}\partial_{\mu}-m_{0}-\sum^{16}_{A=1}\Gamma_{A}v^{A}(x))\psi_{n}(x)=0, \\
v^{S}(x) = \frac{\delta E_{XC}[Q(x)]}{\delta Q^{S}(x)} = \delta m +\alpha(x), \\
v^{V}_{\mu}(x) = A^{ext}_{\mu}(x)+\int d^{3}x'\frac{j_{\mu}(x')}{|{\bf x}-{\bf x}'|}+\frac{\delta E_{XC}[Q(x)]}{\delta Q^{V}_{\mu}(x)}, \\
v^{T}_{\mu\nu}(x) = \frac{\delta E_{XC}[Q(x)]}{\delta Q^{T}_{\mu\nu}(x)}, \\
v^{A}_{\mu}(x) = \frac{\delta E_{XC}[Q(x)]}{\delta Q^{A}_{\mu}(x)}, \\
v^{P}(x) = \frac{\delta E_{XC}[Q(x)]}{\delta Q^{P}(x)}.     
\end{eqnarray}
Here the effective single-particle Kohn-Sham potentials are also presented. 
Now, all of them are time-dependent potentials.

\section{Group theoretical Classification of QED-DMFT Solutions}

\subsection{Group theoretical Classification}

In this section, we give a discussion of group theoretical classification of the QED-DMFT solutions.

In the relativistic theory, we usually treat the Poincar\'{e} group ( 4-translation and the Lorentz group ), charge conjugation, parity ( spatial inversion ) and time-reversal. Under the charge conjugation, $\psi$ and $\bar{\psi}$ are transformed as
\begin{eqnarray}
\psi \to C\bar{\psi}^{T}, \qquad \bar{\psi} \to -\psi^{T}C^{-1},
\end{eqnarray} 
where $T$ denotes the transposition of a matrix, $C\equiv i\gamma^{2}\gamma^{0}$ is the charge conjugation matrix. Then $Q(x)$ is transformed as
\begin{eqnarray}
Q &=& -\langle\psi\bar{\psi}\rangle \to -C\langle\psi\bar{\psi}\rangle ^{T}C^{-1}= CQ^{T}C^{-1} \nonumber \\
&=& Q^{S}I-Q^{V}_{\mu}\gamma^{\mu}-Q^{T}_{\mu\nu}\sigma^{\mu\nu}+Q^{A}_{\mu}\gamma_{5}\gamma^{\mu}+Q^{P}i\gamma_{5}.
\end{eqnarray}
Under the time reversal,
\begin{eqnarray}
\psi(t) \to T\psi(-t), \qquad \bar{\psi}(t) \to \bar{\psi}(-t)T.
\end{eqnarray}
Here $T\equiv i\gamma^{1}\gamma^{3}$. Then, together with the rule of taking the complex conjugate about c-numbers, we obtain
\begin{eqnarray}
Q(t) &=& -\langle\psi(t)\bar{\psi}(t)\rangle \to -T\langle\psi(-t)\bar{\psi}(-t)\rangle ^{*}T =TQ(-t)^{*}T \nonumber \\
&=& Q^{S*}(-t)I+Q^{V*}_{0}(-t)\gamma^{0}-{\bf Q}^{V*}(-t)\cdot\vec{\gamma}+Q^{T*}_{0i}(-t)\sigma^{0i} \nonumber \\
& & -Q^{T*}_{ij}(-t)\sigma^{ij}+Q^{A*}_{0}(-t)\gamma_{5}\gamma^{0}-{\bf Q}^{A*}(-t)\cdot\gamma_{5}\vec{\gamma}-Q^{P*}(-t)i\gamma_{5},
\end{eqnarray}
where $i,j=1,2,3$. $Q^{V}_{0}$ is 0-th component of vector while $Q^{A}_{0}$ is 0-th component of axial vector. ${\bf Q}^{V}$ and ${\bf Q}^{A}$ are space components of vector and axial vector, respectively. Under the spatial inversion,
\begin{eqnarray}
\psi({\bf x}) \to \gamma^{0}\psi(-{\bf x}), \qquad \bar{\psi}({\bf x}) \to \bar{\psi}(-{\bf x})\gamma^{0},
\end{eqnarray}
$Q(x)$ is transformed as
\begin{eqnarray}
Q({\bf x}) &=& -\langle\psi({\bf x})\bar{\psi}({\bf x})\rangle \to -\gamma^{0}\langle\psi(-{\bf x})\bar{\psi}(-{\bf x})\rangle\gamma^{0} =\gamma^{0}Q(-{\bf x})\gamma^{0}\nonumber \\
&=& Q^{S}(-{\bf x})I+Q^{V}_{0}(-{\bf x})\gamma^{0}-{\bf Q}^{V}(-{\bf x})\cdot\vec{\gamma}-Q^{T}_{0i}(-{\bf x})\sigma^{0i}\nonumber \\
& & +Q^{T}_{ij}(-{\bf x})\sigma^{ij}-Q^{A}_{0}(-{\bf x})\gamma_{5}\gamma^{0}+{\bf Q}^{A}(-{\bf x})\cdot\gamma_{5}\vec{\gamma}-Q^{P}(-{\bf x})i\gamma_{5}. 
\end{eqnarray}

The QED-DMFT solutions can be group theoretically classified into several types. To consider this problem, we determine the symmetry group of a system. In the atomic or molecular systems, the translation invariance is broken. In the case of an atom, only $O(3)$ rotation remains in the Lorentz group ( In the case of a molecule, $O(3)$ is replaced by the point group. ). Under the $O(3)$ rotational symmetry, we expand the $Q^{S}$, $Q^{P}$, $Q^{V}_{0}$ and $Q^{A}_{0}$ by the scalar spherical harmonics, while we expand the ${\bf Q}^{V}$, ${\bf Q}^{A}$ and $Q^{T}_{\mu\nu}$ by the vector spherical harmonics~[36]. It is clear from (41) and (43), the behavior of each type of the density matrix given in (12) under the spatial inversion and time reversal depends not only on the structure of the $\gamma$-matrix, but also on the angular momentum of the spherical harmonics. Let us consider the case of an atom. We treat the group $G=O(3)\times P\times T$. We introduce the subgroup of $G$ as
\begin{eqnarray}
& & O(3)\times P\times T, \quad O(3)\times P, \quad O(3)\times T, \nonumber \\
& & P\times T, \quad O(3), \quad P, \quad T, \quad 1. 
\end{eqnarray}
For example, $O(3)\times P\times T$-invariant solution is given as
\begin{eqnarray}
Q^{S*}=Q^{S},\quad Q^{V*}_{0}=Q^{V}_{0},\quad {\rm others}=0.
\end{eqnarray}
There is no case for $O(3)\times P$-invariant solution.
$O(3)\times T$-invariant solution is given as
\begin{eqnarray}
Q^{S*}=Q^{S},\quad Q^{V*}_{0}=Q^{V}_{0},\quad Q^{A*}_{0}=Q^{A}_{0},\quad {\rm others}=0.
\end{eqnarray}
$O(3)$-invariant solution is given as
\begin{eqnarray}
Q^{S*}=Q^{S},\quad Q^{V*}_{0}=Q^{V}_{0},\quad Q^{A*}_{0}=Q^{A}_{0},\quad Q^{P*}=Q^{P},\quad {\rm others}=0.
\end{eqnarray}
Due to the $O(3)$ rotational invariance, each density matrix can only have an s-wave component in all cases (45)$\sim$(47). All solutions given above are the cases for closed shell states. We may have magnetic QED-DMFT solutions for systems under discussions, if time reversal symmetry is broken ( for example, the case of (47) ). 

Under the presence of the vectorial density matrices like the ${\bf Q}^{V}$, ${\bf Q}^{A}$ and $Q^{T}_{\mu\nu}$, the $O(3)$ rotational symmetry will be broken. For $P\times T$-, $P$-, $T$- and $1$- (no symmetry) invariant solutions,
simple classification is impossible, because there are various possibilities
of the angular momentum dependences of $Q$. This situation demands us futher investigation in more detail in the future.
It must be noted that we have to solve the Dirac-Kohn-Sham equation to find which type of the density matrices ( and solutions ) will be realized.

\subsection{Nonrelativistic and Ultrarelativistic Limits}

To consider the nonrelativistic and ultrarelativistic limits, we take the standard representation. The four component bispinor $\psi$ is partitioned as~[36]
\begin{equation}
\psi=\left(
\begin{array}{c}
\phi \\
\chi
\end{array}
\right),
\end{equation}
where $\phi$ is the large component and $\chi$ is the small component. Then we obtain 
\begin{eqnarray}
Q^{S} &=& \frac{1}{4}{\rm tr} Q = \frac{1}{4}\langle\bar{\psi}\psi\rangle =\frac{1}{4}\langle\phi^{\dagger}\phi-\chi^{\dagger}\chi\rangle, \\
Q^{V}_{0} &=& \frac{1}{4}{\rm tr} \gamma^{0}Q = \frac{1}{4}\langle\bar{\psi}\gamma^{0}\psi\rangle =\frac{1}{4}\langle\phi^{\dagger}\phi+\chi^{\dagger}\chi\rangle, \\
{\bf Q}^{V} &=& \frac{1}{4}{\rm tr} \vec{\gamma}Q = \frac{1}{4}\langle\bar{\psi}\vec{\gamma}\psi\rangle =\frac{1}{4}\langle\chi^{\dagger}\vec{\sigma}\phi+\phi^{\dagger}\vec{\sigma}\chi\rangle, \\
{\bf Q}^{T}_{(V)} &=& \frac{1}{4}{\rm tr} \sigma^{0i}Q = \frac{1}{4}\langle\bar{\psi}i\gamma^{0}\vec{\gamma}\psi\rangle =\frac{i}{4}\langle-\chi^{\dagger}\vec{\sigma}\phi+\phi^{\dagger}\vec{\sigma}\chi\rangle, \\
{\bf Q}^{T}_{(A)} &=& \frac{1}{4}{\rm tr} \sigma^{ij}Q = \frac{1}{4}\langle\bar{\psi}\gamma_{5}\gamma^{0}\vec{\gamma}\psi\rangle =\frac{1}{4}\langle\phi^{\dagger}\vec{\sigma}\phi-\chi^{\dagger}\vec{\sigma}\chi\rangle, \\
Q^{A}_{0} &=& \frac{1}{4}{\rm tr} \gamma_{5}\gamma^{0}Q = \frac{1}{4}\langle\bar{\psi}\gamma_{5}\gamma^{0}\psi\rangle =-\frac{1}{4}\langle\chi^{\dagger}\phi+\phi^{\dagger}\chi\rangle, \\
{\bf Q}^{A} &=& \frac{1}{4}{\rm tr} \gamma_{5}\vec{\gamma}Q = \frac{1}{4}\langle\bar{\psi}\gamma_{5}\vec{\gamma}\psi\rangle = -\frac{1}{4}\langle\phi^{\dagger}\vec{\sigma}\phi+\chi^{\dagger}\vec{\sigma}\chi\rangle, \\
Q^{P} &=& \frac{1}{4}{\rm tr} i\gamma_{5} Q = \frac{1}{4}\langle\bar{\psi}i\gamma_{5}\psi\rangle =\frac{1}{4}\langle-i\chi^{\dagger}\phi+i\phi^{\dagger}\chi\rangle. 
\end{eqnarray}
Here, ${\bf Q}^{T}_{(V)}$ denotes vector-like components of $Q^{T}_{\mu\nu}$, while ${\bf Q}^{T}_{(A)}$ denotes axial-vector-like components of $Q^{T}_{\mu\nu}$. Therefore, at the nonrelativistic limit ( $\chi$ will vanish ), the $Q^{S}$ and $Q^{V}_{0}$ coincide with each other and they give $\frac{1}{4}\phi^{\dagger}\phi$, while the ${\bf Q}^{T}_{(A)}$ gives $\frac{1}{4}\phi^{\dagger}\vec{\sigma}\phi$ and the ${\bf Q}^{A}$ gives $-\frac{1}{4}\phi^{\dagger}\vec{\sigma}\phi$. The ${\bf Q}^{V}$, ${\bf Q}^{T}_{(V)}$, $Q^{A}_{0}$ and $Q^{P}$ vanish at the nonrelativistic limit. $\phi^{\dagger}\phi$ corresponds to the usual density of nonrelativistic theory, while $\phi^{\dagger}\vec{\sigma}\phi$ corresponds to the spin density of nonrelativistic theory. Futhermore, at the ultrarelativistic limit ( $\phi$ and $\chi$ coincide with each other ), $Q^{V}_{0}$, ${\bf Q}^{V}$, $Q^{A}_{0}$ and ${\bf Q}^{A}$ remain while others vanish. Therefore, for a system under strong relativistic effects, $Q^{V}_{\mu}$ and $Q^{A}_{\mu}$ may give important contributions to QED-DMFT solutions.

\section{Discussion}

In this paper, we have investigated the QED-DMFT and the group theoretical classification of its solutions. We have introduced the one-particle local density matrix $Q(x)$, and have decomposed by using $\gamma$-matrices. We have discussed the transformation properties of the matrix under group operations, and its nonrelativistic limit. 

Now, we discuss the application and extension of our theory. 

In the context of the electronic structure of atoms, there are three effects: The electron correlation, the relativistic effect and the QED effect. The electronic structure of atoms is determined by the relation of these three factors. The electron correlation depends on the electron numbers and can be treated by several many-body techniques. The relativistic effect ( kinematic effect, spin-orbit, Darwin, mass-velocity, etc. ) becomes large with increasing the atomic number. On the other hand, the case to describe the inner core electrons of heavy elements or, electrons of highly ionized heavy elements such as lithium-like uranium, the QED effect ( mainly originate in negative energy state ) can not be neglected and we must take the Dirac sea into account. In the case of heavy atoms, as the ionicity becomes high and the electron number decrease, the many-body effect becomes small and the QED effect becomes large. In principle, QED-DMFT can exactly estimate the three effects discussed above. Therefore we propose that QED-DMFT should be applied to the cases where both the many-body effects and the QED effects can not be neglected. Practically, we need a well-behaved exchange-correlation functional which contains both the correlation energy and QED effect satisfactorily. Heavy elements in middle level of ionicity should be one of the subject for our theory. The calculation of electromagnetic properties, for example, g-factors, hyperfine interactions, nuclear magnetic resonance shielding constants, are interesting. The collision of two uranium atoms is also one of the interesting subject of QED-TDDMFT. QED-TDDMFT should be useful to describe excitations, susceptibilities and collective modes of systems. Recently, experiments of x-ray irradiation to cluster plasma are performed, and verious new phenomena were studied~[37]. The cluster of heavy ions under middle revel of ionicity is now an interesting subject in this area. The importance of the relativistic and QED effects is discussed in such objects. 

It is a famous fact that, the available exchange-correlation functionals only describe the dynamic correlations reliably, and they fails for systems with strong static correlations. Thus, usual Kohn-Sham scheme often breaks down in the case of near degenerate ground states. In the usual method of quantum chemistry, the static correlation should be described by using multi-determinantal wavefunctions. The CASSCF is one of the methods to take into account the static correlation effect correctly. Recently, the CAS-DFT method was proposed~[38]. In this method, the dynamic correlation is considered by usual DFT scheme, while static correlation is handled by CAS-type wavefunction. The energy functional is given as
\begin{eqnarray}
E[\rho] &=& \min_{\rho\to N}\Bigl\{ F[\rho]+\int d^{3}x\rho({\bf x})v({\bf x})\Bigr\}, \\ 
F[\rho] &=& F^{CAS}_{KS}[\rho]+E^{CAS}_{C}[\rho], \\
F^{CAS}_{KS}[\rho] &=& \min_{\Psi^{CAS}\to \rho}\langle\Psi^{CAS}|\hat{T}+\hat{V}_{ee}|\Psi^{CAS}\rangle,  
\end{eqnarray}
where $\hat{T}$ and $\hat{V}_{ee}$ are the kinetic energy and electron-electron interaction, respectively. $\Psi^{CAS}$ is a CAS trial wavefunction. In the scheme given above, exchange energy and static correlation energy are covered by $F^{CAS}_{KS}$, while $E^{CAS}_{C}$ includes the dynamic correlation energy only. CAS-DFT in nonrelativistic case gives accurate results~[38]. Thus we propose that this method can be extended to the relativistic theory, and should become usuful tool to consider near degeneracy effect in relativistic DFT framework. For this aim, we introduce the density matrix which is defined as
\begin{equation}
\left(
\begin{array}{cc}
\rho^{++}_{ij} & \rho^{+-}_{ij} \\
\rho^{-+}_{ij} & \rho^{--}_{ij}
\end{array}
\right)= \left(
\begin{array}{cc}
\langle\Phi_{GS}|\hat{a}^{\dagger}_{j}\hat{a}_{i}|\Phi_{GS}\rangle &
\langle\Phi_{GS}|\hat{b}_{j}\hat{a}_{i}|\Phi_{GS}\rangle  \\
\langle\Phi_{GS}|\hat{a}^{\dagger}_{j}\hat{b}^{\dagger}_{i}|\Phi_{GS}\rangle &
\langle\Phi_{GS}|\hat{b}_{j}\hat{b}^{\dagger}_{i}|\Phi_{GS}\rangle 
\end{array}
\right),
\end{equation}
Then we diagonalize the density matrix, we obtain the natural orbital given by four-component bispinor $\eta^{(\pm)}_{i}$ and $n_{i}$ as its occupation number. By using the occupation number $n_{i}$, we can select the active space for treatments of the near degeneracy effects~[39]. The natural orbital analysis of a resulted solution is useful to elucidate the type and magnitude of broken symmetry.

It is also interesting for us to use the information entropy to measure the electron correlation in various systems~[40]. For example, Shannon entropy 
\begin{eqnarray}
S^{Shannon}(\rho) &=& -\int d^{3}x \rho({\bf x})\ln\rho({\bf x}), 
\end{eqnarray}
or Jaynes entropy
\begin{eqnarray}
S^{Jaynes} &=& -\sum_{i}n_{i}\ln n_{i}, 
\end{eqnarray}
might be useful tools. The electron correlation energy can be estimated based on the Collins's conjecture:
\begin{eqnarray}
E_{corr} &=& -k\sum_{i}n_{i}\ln n_{i} \propto S^{Jaynes}.
\end{eqnarray}
Therefore, by using these entropies, we propose to estimate the strength of the electron correlation in various atoms under various ionicities with QED-DMFT. 
The relations between various theories proposed in this paper are depicted in Fig. 2.

\acknowledgments

The authors wish to thank colleagues of the quantum chemistry laboratory at Osaka university, for their aid. One of the authors (T.O) would like to express his gratitude sincerely to Professor Hiroshi Toki (RCNP, Osaka univ.), for his invariable kindhearted care.

\begin{figure}

\caption{The relations between various DFT. Meanings of abbreviations are given in text.}

\caption{The relations between various theories proposed in text.}

\end{figure}

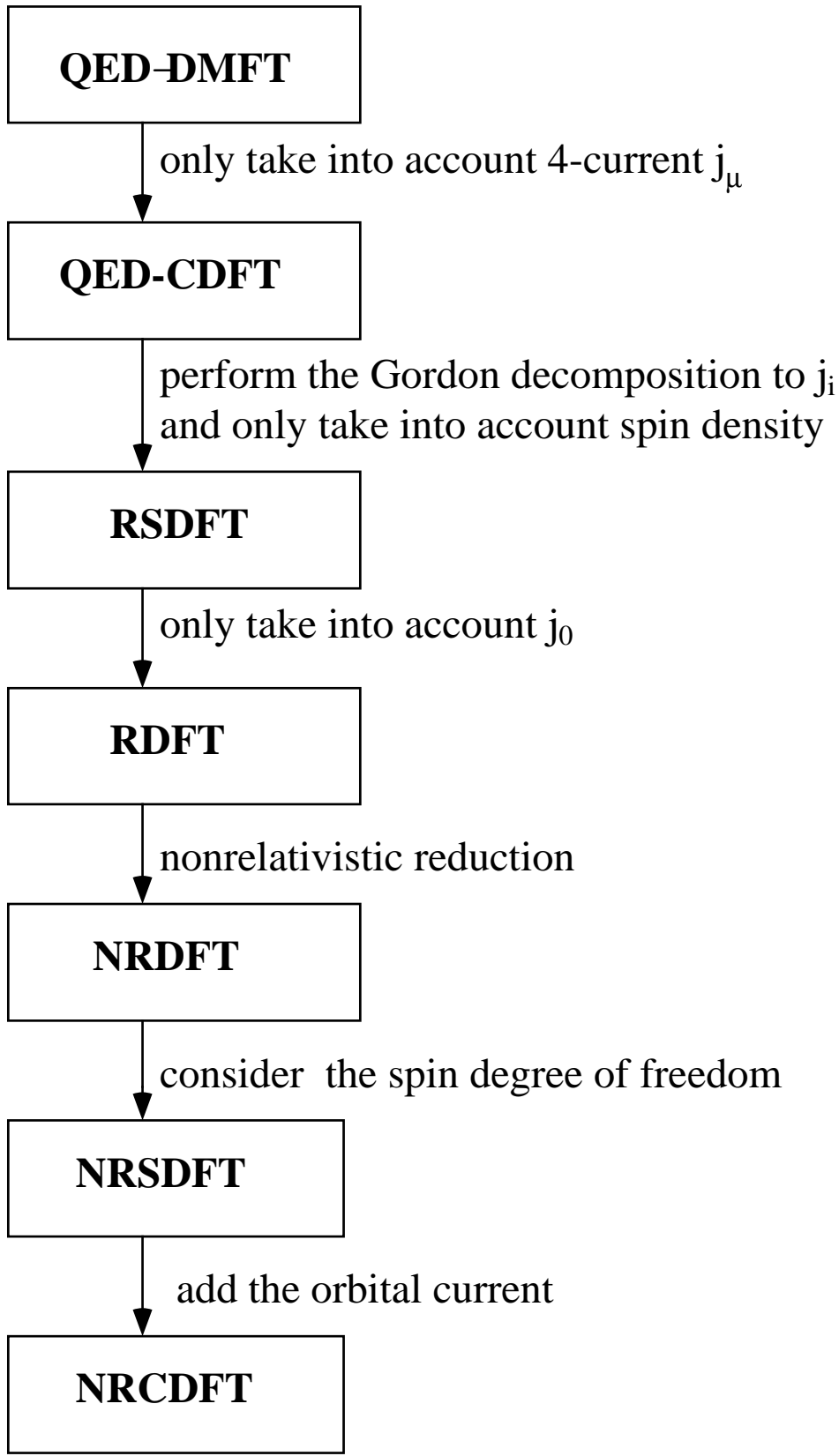
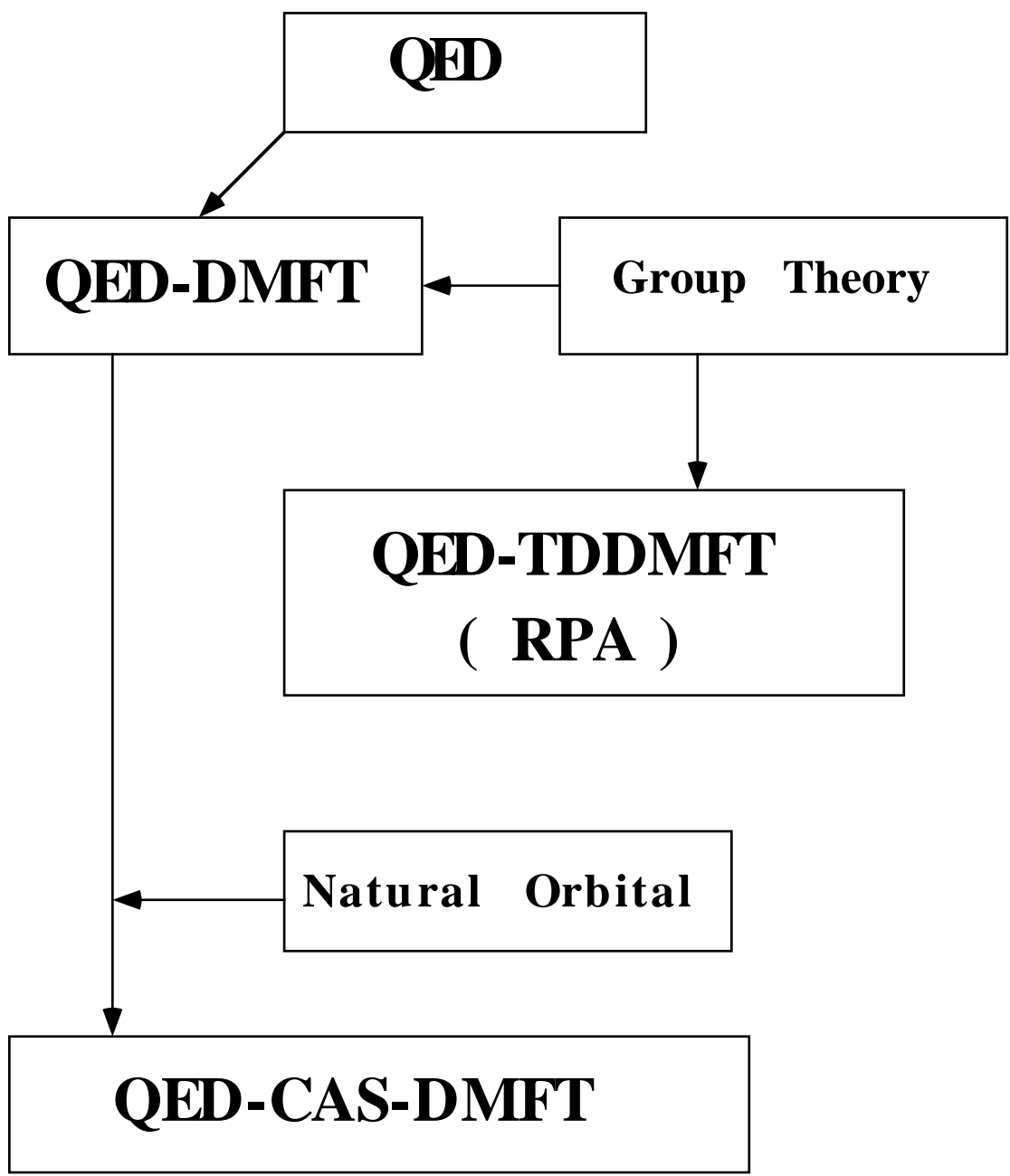
\end{document}